\documentclass[11pt]{article}
\usepackage{amsmath}
\usepackage{amssymb}
\usepackage{amsfonts}
\usepackage{amscd}
\usepackage{accents}
\usepackage{bbm} 
\usepackage{pst-all}
\usepackage{epsfig}
\date{\today}
\newcommand{\bmat}{\left(\begin{array}}
\newcommand{\emat}{\end{array}\right)}
\newcommand{\be}{\begin{equation}}
\newcommand{\ee}{\end{equation}}
\newcommand{\ba}{\begin{eqnarray}}
\newcommand{\ea}{\end{eqnarray}}

\def\lsim{\raise0.3ex\hbox{$\;<$\kern-0.75em\raise-1.1ex\hbox{$\sim\;$}}}
\def\gsim{\raise0.3ex\hbox{$\;>$\kern-0.75em\raise-1.1ex\hbox{$\sim\;$}}}

\def\be{\beta}

\begin{document}
\vspace*{-.6in} \thispagestyle{empty}
\begin{flushright}
DESY 12-028
\end{flushright}
\baselineskip = 20pt

\vspace{.5in} {\Large
\begin{center}
{\bf On Stability of the Electroweak Vacuum and the  Higgs Portal }

\end{center}}

\vspace{.5in}

\begin{center}
{\bf  Oleg Lebedev    }  \\

\vspace{.5in}

 \emph{DESY Theory Group, 
Notkestrasse 85, D-22607 Hamburg, Germany
 }
\end{center}

\vspace{.5in}

\begin{abstract}
In the Standard Model (SM),  the Higgs mass  around 125 GeV implies
that  the electroweak vacuum is metastable
since the quartic Higgs coupling turns negative at  high energies. 
I point out that a tiny mixing of the Higgs with a 
heavy singlet can make the electroweak vacuum completely stable.
This is due to a tree level correction to the Higgs mass--coupling relation,
which survives in the zero--mixing/heavy--singlet  limit.  
Such a  situation is experimentally indistinguishable from the SM,
unless the Higgs self--coupling can be measured. 
As a result, Higgs inflation and its variants can
still be viable.
\end{abstract}

\noindent

\newpage

\section{Introduction}

The Higgs sector of the Standard Model (SM)  enjoys a special status. The Higgs bilinear
$H^\dagger H$ is the only gauge and Lorentz invariant dimension--2 operator that
can be constructed out of the SM fields. This operator can couple to the ``hidden'' sector
at the renormalizable level \cite{Silveira:1985rk,UM-P-91-54},
\begin{equation} 
\Delta {\cal L} = c~ H^\dagger H S^2 \;,
\end{equation}
where $S$ is an SM singlet. More general allowed couplings include, for example, a coupling to  
massive vectors \cite{Kanemura:2010sh,Lebedev:2011iq}
and a scalar curvature  \cite{Bezrukov:2007ep}, 
which could be relevant to dark matter and inflation, respectively. 
In the case of a scalar Higgs portal, the phenomenology depends crucially on whether
or not $S$ develops a vacuum expectation value (VEV). If it does, the Higgs boson mixes
with the singlet  \cite{Schabinger:2005ei,Patt:2006fw}, 
otherwise $S$ becomes  a good dak matter candidate  
\cite{Silveira:1985rk,McDonald:1993ex,Burgess:2000yq}.
Here we focus on the first possibility. A significant  Higgs--singlet mixing can be 
probed at the LHC 
\cite{Barger:2007im,Englert:2011yb} by measuring production cross sections for 
the Higgs--like states, whereas 
the small mixing ($<$ 10\%) case is much more challenging.

In this paper, we explore some consequences of an unobservably small Higgs--singlet   mixing. 
We find, in particular, that such a mixing together with tiny singlet couplings
 can stabilize the electroweak (EW) vacuum,
which otherwise appears metastable \cite{EliasMiro:2011aa,Espinosa:2007qp}.   
Furthermore, it can  lead to a significant (${\cal O}(1)$) tree--level modification of   
 the Higgs self--coupling, which can be measured at colliders. 
Finally, although Higgs inflation within the SM is disfavored by 
the tentative  125 GeV Higgs signal  seen at the LHC \cite{Collaboration:2012si,Chatrchyan:2012tx},
such a possibility remains open in our framework.

\section{Higgs portal potential and stability }

We start by reviewing properties  of the Higgs portal potential 
following Ref.~\cite{Lebedev:2011aq}. 
Related work can be found in 
\cite{Bowen:2007ia,Djouadi:2011aa,Mambrini:2011ik}.

\subsection{Relevant formulae}

The scalar potential  in the unitary gauge $H^T=(0,h/\sqrt{2})$
is given by
\begin{equation}
V= {1\over 4} \lambda_h h^4 + {1\over 4}\lambda_{hs} s^2 h^2 + {1\over 4}\lambda_s s^4 + 
{1\over 2} m_h^2 h^2 + {1\over 2} m_s^2 s^2 \;. \label{V}
\end{equation}
Here $h$ and $s$ are real fields; 
$\lambda_i$ are the quartic  couplings and $m_i^2$ are mass parameters.
In what follows, we will be interested in the case when both $h$ and $s$ develop
non--zero VEVs. Denoting  $\langle h \rangle =v, \langle s \rangle =u$,
extremization of the low energy scalar potential  (\ref{V})   requires
\begin{eqnarray}
&& v^2 = 2~ {   \lambda_{hs} m_s^2 - 2 \lambda_s m_h^2 \over
4 \lambda_s \lambda_h -   \lambda_{hs}^2  } \;, \nonumber\\
&& u^2 = 2~ {   \lambda_{hs} m_h^2 - 2 \lambda_h m_s^2 \over
4 \lambda_s \lambda_h -   \lambda_{hs}^2  } \;.  \label{VEV}
\end{eqnarray}
The diagonal matrix elements of the Hessian at this point are
$2 \lambda_s u^2$ and  $2 \lambda_h v^2$, while its determinant is
$ (4 \lambda_s \lambda_h -   \lambda_{hs}^2)v^2 u^2  $. Then,
the extremum is a local minimum  if 
\begin{eqnarray}
&& \lambda_{hs} m_h^2 - 2 \lambda_h m_s^2 > 0 \;,\nonumber\\
&& \lambda_{hs} m_s^2 - 2 \lambda_s m_h^2 > 0 \;,\nonumber\\
&& 4 \lambda_s \lambda_h -   \lambda_{hs}^2 >0 \;. \label{ewbreaking}
\end{eqnarray}
In this case, the mass squared eigenvalues are
\begin{equation}
m_{1,2}^2= \lambda_h v^2 + \lambda_s u^2 \mp 
\sqrt{(\lambda_s u^2 - \lambda_h v^2)^2 + \lambda_{hs}^2 u^2 v^2 }
\label{eigenvalues}
\end{equation}
with the mixing angle $\theta$ given by
\begin{equation}
\tan 2 \theta = {\lambda_{hs} u v  \over \lambda_h v^2 - \lambda_s u^2} \;.
\label{tan}
\end{equation}
Following the convention of \cite{Lebedev:2011aq}, the mixing angle is defined by 
\begin{equation}
O^T~ M^2 ~ O = {\rm diag}(m_1^2, m_2^2) ~~,~~ O=\left( 
\begin{matrix} 
\cos\theta & \sin\theta \\
-\sin\theta & \cos\theta  
\end{matrix} 
\right) \;,
\end{equation}
where $M^2$ is a 2$\times$2 mass squared matrix.
The range of $\theta$ is related to the ordering of the eigenvalues through
$ {\rm sign} (m_1^2-m_2^2 ) = {\rm sign} (\lambda_s u^2 - \lambda_h v^2)~
 {\rm sign} (\cos 2 \theta) $ and we take $m_1$ to be the smaller eigenvalue.
The mass eigenstates are
\begin{eqnarray}
 H_1 &=&  s \cos\theta  - h \sin\theta  \;, \nonumber\\
 H_2 &=&  s \sin\theta  + h \cos\theta  \;. 
\end{eqnarray}
Note that the lighter   mass eigenstate  $H_1$  is 
``Higgs--like'' for $\lambda_s u^2 > \lambda_h v^2$ and 
``singlet--like'' otherwise. 
The former case corresponds to $\vert \theta\vert >\pi/4$.

\subsection{Large singlet VEV limit}

In the limit $u \gg v$, we have 
\begin{eqnarray}
&& m_1^2 \simeq 2 \left(   \lambda_h -{\lambda_{hs}^2 \over 4 \lambda_s }      \right)~v^2 \;,
\nonumber\\
&& m_2^2 \simeq 2 \lambda_s u^2 + {\lambda_{hs}^2 \over 2 \lambda_s }~v^2 \;,
\nonumber\\
&& \tan 2\theta \simeq - {\lambda_{hs} v \over \lambda_s u}\;,
\end{eqnarray}
where the neglected terms are suppressed by $v^2/u^2$. 
As $u$ increases,  $\tan 2\theta \rightarrow 0$ and the singlet--Higgs mixing 
approaches zero. The light eigenstate $H_1$ is almost a pure Higgs, yet 
its mass squared is not $2 \lambda_h v^2$ as in the SM,  but receives
a finite negative correction $ -\lambda_{hs}^2 v^2/  (2 \lambda_s)$. 
Therefore, a given Higgs mass corresponds to a $larger$ $\lambda_h$ than what
it would be in the Standard Model.

A 125 GeV Higgs can then be obtained for 
various $\lambda_h$ up to order one as long as  
\begin{equation}
\lambda_h -{\lambda_{hs}^2 \over 4 \lambda_s }   \simeq 0.13 \;.
\end{equation}
As a result,  all the couplings   can remain positive at all scales,
ensuring stability of the potential. This can be achieved with tiny
$\lambda_{hs}$ and $\lambda_s$ as long as $\lambda_{hs}^2/\lambda_s $
is significant.

As one increases  $u \gg v$ (while keeping both states as relevant degrees of
freedom at a given energy scale), the effect on the light eigenvalue remains finite.
The SM result is not recovered 
due to the large cross term $\lambda_{hs} u^2 v^2$, although the light state
is (almost) exactly SM--like. The cross term generates a 
leading order  ${\cal O}(u^2 v^2)$
correction to the determinant of the mass squared matrix, resulting in  a 
finite shift in $m_1$.

Rewriting
\begin{eqnarray}
&& m_h^2 = -{1\over 2} \lambda_{hs} u^2 -\lambda_h v^2 \;, \nonumber\\
&& m_s^2 = -{1\over 2} \lambda_{hs} v^2 -\lambda_s u^2 \;,
\end{eqnarray}
one finds that the $u \rightarrow \infty$ limit means that 
both $m_s^2$ and $m_h^2$ increase in magnitude   indefinitely, 
while $m_s^2/m_h^2 \rightarrow 2\lambda_s /\lambda_{hs}$.  
The hierarchy between $v$ and $u$ is then equivalent to tuning 
of $m_s^2/m_h^2$ to this value. Note also that the vacuum energy
is negative and scales like $- u^4$.

Since the singlet--Higgs mixing can be very small, 
this scenario is essentially indistinguishable from the SM apart from
the Higgs quartic coupling. 
Suppose that both mass eigenstates are sufficiently light so that they
are degrees of freedom of our TeV--scale effective theory,
\begin{equation}
m_1 \ll m_2 \sim E \;,
\end{equation}
with $E \sim {\cal O}$(TeV).
Given the Higgs mass $m_{\rm Higgs}$ (which we also
identify with $m_1$), the quartic couplings are given by
\begin{eqnarray}
 \lambda_h \big\vert_{\rm SM} &=& {m_{\rm Higgs}^2 \over 2 v^2} \;, \nonumber\\
 \lambda_h \big\vert_{\rm SM+ singlet}&=& {m_{\rm Higgs}^2 \over 2 v^2} + 
{\lambda_{hs}^2 \over 4 \lambda_s} \;.
\end{eqnarray}
For a TeV--scale singlet, the correction to the quartic Higgs coupling can be 
of order 100\%, whereas the gauge--Higgs coupling is modified by less than 10\%.
An ILC \cite{Gounaris:1979px}   
or high--luminosity LHC \cite{Baur:2003gp}  
study of the Higgs self--coupling would 
then be able to reveal the presence of the singlet.

If the singlet is very heavy,
\begin{equation}
m_2 \gg {\rm TeV} \;,
\end{equation} 
the effective low energy theory
contains just the Higgs doublet and the Higgs mass--coupling relation is not
affected \cite{EliasMiro:2012ay}. The quartic coupling receives the singlet--induced
contribution above  the  singlet threshold, which has an analogous stabilizing effect.
In this case, the mixing $ \sim v/u$ can be as small as $10^{-7}$ with the 
constraint coming from ``activating'' the singlet below the SM instability scale.

\subsection{Numerical example}

The quartic couplings at high energies are determined by the following 1--loop
renormalization group equations  (see e.g. \cite{singletinflation})
\begin{eqnarray}
16\pi^2 {d \lambda_h \over dt}&=& 24 \lambda_h^2 -6 y_t^4 +{3\over 8} \Bigl( 
2 g^4 + (g^2 + g^{\prime 2})^2 \Bigr) \nonumber\\
&+& (-9 g^2 -3 g^{\prime 2}+12 y_t^2) \lambda_h + {1\over 2} \lambda_{hs}^2 \;, \nonumber\\
16\pi^2 {d \lambda_{hs} \over dt} &=& 4 \lambda_{hs}^2 + 12 \lambda_h \lambda_{hs}
-{3\over 2} (3 g^2 + g^{\prime 2}) \lambda_{hs}  \nonumber\\
&+& 6 y_t^2  \lambda_{hs} +
6 \lambda_s  \lambda_{hs} \;, \nonumber\\
 16\pi^2 {d \lambda_{s} \over dt} &=& 2 \lambda_{hs}^2 + 18 \lambda_s^2 \;,
\end{eqnarray}
where $t = \ln (\mu/ m_t)$ with $\mu$ being the energy scale.
The effect of the heavy singlet threshold can be included by inserting appropriate
 Heaviside $\theta$--functions.
The RG equations for the gauge and the top Yukawa couplings are given by the usual SM
expressions.
 The low energy input values for these couplings are 
$g(m_t)=0.64 , g'(m_t)=0.35 , g_3(m_t)=1.16$, while for the top Yukawa coupling we use
its running value at $m_t$, $y_t (m_t)=0.93$ \cite{Langenfeld:2010aj}.
A two--loop SM running of $\lambda_h$ as well as finite corrections are  considered
in a recent analysis \cite{EliasMiro:2011aa} (see \cite{Holthausen:2011aa} for a discussion). 
Within  the uncertainties of $m_t$,
our 1--loop SM results are consistent with theirs.

\begin{figure}
\hspace*{2.5cm}
\includegraphics[width=10.0cm]{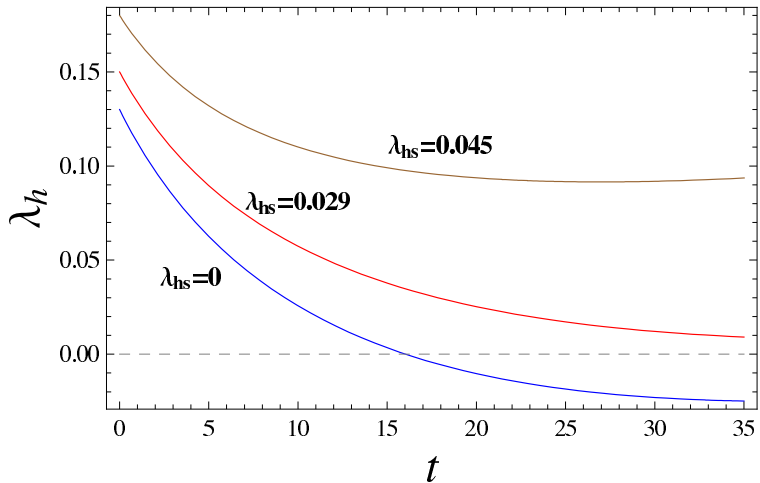}
\vspace*{0.3cm}
 \caption{\footnotesize Higgs quartic coupling evolution with energy for various $\lambda_{hs}$
 and $\lambda_s=0.01$ at $m_t$.  The Higgs mass is fixed at
approximately 125 GeV. }
\label{f1} 
\end{figure}

Figure \ref{f1} displays a few possible choices of low energy couplings consistent with
$m_h \simeq 125$ GeV and the resulting evolution of $\lambda_h$ with energy.
We have chosen small couplings $\sim 10^{-2}$ in which case the 
singlet loop contribution to $\lambda_h$ is suppressed, as is the singlet threshold effect
(for definiteness, we have set the singlet mass to be sub--TeV).
For 
$ \lambda_{hs}^2 / (4 \lambda_s)  $ greater than about 0.015, the Higgs quartic coupling is positive at 
all scales. As seen from the RG equations, $\lambda_s$  remains positive if it is positive at low 
energies. This also applies to $\lambda_{hs}$ as its evolution is dominated by the top--quark
contribution. The potential is therefore positive--definite at high energies and the EW 
vacuum is stable.  Note that this result can be achieved with tiny $ \lambda_{hs}$ and 
$\lambda_{s}$ since what matters is $ \lambda_{hs}^2 / (4 \lambda_s)  $.

Analogous examples can be constructed  for $\lambda_{hs}<0$ as well.
In that case, one must obey the additional constraint 
$    4 \lambda_s \lambda_h -   \lambda_{hs}^2 >0    $ at all scales.
Even though this quantity is positive in the low energy limit, 
it may turn negative at higher energies signifying the existence of a 
run--away direction.

This  illustrates how the electroweak vacuum can be stabilized
by invoking a heavy scalar which has  practically no observable effects other
than changing the Higgs self--coupling. 
In that case, the bound on the reheating temperature \cite{EliasMiro:2011aa}
relaxes and, furthermore, the Higgs field can be responsible for inflation \cite{Bezrukov:2007ep}.

One may also stabilize the Higgs potential  entirely by loop effects involving a relatively light singlet
with a zero VEV   \cite{Gonderinger:2009jp,Kadastik:2011aa,Chen:2012fa}.
As seen from the RG equations, the singlet makes a positive contribution to $\lambda_h$, so  
 a substantial cross coupling $\lambda_{hs} \sim 0.4$ would stabilize the electroweak vacuum. 
Although viable, this mechanism is quite constraining and the options for Higgs--like inflation
would be  limited. In contrast, the tree--level mechanism described above works even 
for very small couplings and a  heavy singlet 
providing ample opportunities for model building.

\section{Rescuing  Higgs inflation}

Scalar fields with the usual quartic potentials can drive inflation if they possess large
couplings $\xi_i \gg 1$  to scalar curvature $R$ \cite{Bezrukov:2007ep,Bezrukov:2010jz}. 
Below we summarize  basic ideas of
Higgs portal inflation, closely following  Ref.~\cite{Lebedev:2011aq}.
Variants of this scenario which include dark matter considerations
have appeared in 
\cite{singletinflation,Clark:2009dc}.

\subsection{The setup}

The relevant Jordan frame Lagrangian in the unitary gauge is 
\begin{equation}
{\cal L}/ \sqrt{-g} =  - {1\over 2} M^2_{\rm Pl} R - {1\over 2}   \xi_h h^2 R - {1\over 2}   \xi_s s^2 R    
+ {1\over 2} (\partial_\mu h)^2 + {1\over 2} (\partial_\mu s)^2
- V                
\end{equation}
with  $\xi_{h,s} >0$. The scalar couplings to curvature can be eliminated by a conformal
transformation
\begin{equation}
\tilde g_{\mu\nu}=\Omega^2 g_{\mu\nu} ~~,~~
\Omega^2= 1 + { \xi_h h^2 +  \xi_s s^2 \over M_{\rm Pl}^2} \;, \label{Omega}
\end{equation}
which brings us to the Einstein frame. Let us set  $M_{\rm Pl}$ to 1 and consider
the limit 
\begin{equation}
\xi_h h^2 +  \xi_s s^2 \gg  1 \;. 
\end{equation}
At  $ \xi \equiv \xi_h  +\xi_s \gg 1$, the kinetic terms and the potential for new variables
\begin{eqnarray}
&& \chi =  \sqrt{3\over 2}  ~ \log (\xi_h h^2 +  \xi_s s^2) \;, \nonumber\\
&& \tau = {h\over s} \; 
\end{eqnarray}
are given by 
\begin{equation}
{\cal L}_{\rm kin}= {1\over 2 } (\partial_\mu \chi)^2 +
{1\over 2} { \xi_h^2 \tau^2 +\xi_s^2 \over (\xi_h \tau^2 +\xi_s)^3 }(\partial_\mu \tau)^2 \;
\end{equation}
and 
\begin{equation}
U= {\lambda_h \tau^4 + \lambda_{hs} \tau^2 +\lambda_s  \over
 4 (\xi_h \tau^2 +\xi_s)^2} \;,  \label{U}
\end{equation}
respectively. Note that at large $\chi$, the potential is independent of $\chi$ which 
allows for slow--roll inflation. The composition of the inflaton depends on the value
of $\tau$ at the minimum of the potential. The minima of $U$ are classified as follows
 \begin{eqnarray}
&& (1)~2 \lambda_h \xi_s - \lambda_{hs} \xi_h >0~,~
   2 \lambda_s \xi_h - \lambda_{hs} \xi_s >0~,~~~~\tau = \sqrt{ 
 2 \lambda_s \xi_h - \lambda_{hs} \xi_s  \over 
  2 \lambda_h \xi_s - \lambda_{hs} \xi_h   }   \;, \nonumber\\
&& (2)~2 \lambda_h \xi_s - \lambda_{hs} \xi_h >0~,~
   2 \lambda_s \xi_h - \lambda_{hs} \xi_s <0~,~~~~\tau=0  \;, \nonumber\\
&& (3)~2 \lambda_h \xi_s - \lambda_{hs} \xi_h <0~,~
   2 \lambda_s \xi_h - \lambda_{hs} \xi_s >0~,~~~~\tau=\infty  \;, \nonumber\\
&& (4)~2 \lambda_h \xi_s - \lambda_{hs} \xi_h <0~,~
   2 \lambda_s \xi_h - \lambda_{hs} \xi_s <0~,~~~~\tau=0,\infty  \;. \label{taumin}
\end{eqnarray}
In all of these minima, the $\tau$ field is heavy ($m\sim 1/\sqrt{\xi}$ in Planck units)
and can be integrated out, leaving $\chi$ as the only dynamical variable during
inflation. If we are to retain the subleading $M_{\rm Pl}^2/(\xi_h h^2 +  \xi_s s^2)$ term
in $\Omega^2$, the potential for $\chi$ becomes 
\begin{equation}
 U(\chi)= {\lambda_{\rm eff}  \over 4\xi_h^2}~
 \Bigl(    1+ {\rm exp}\left( -{2\chi\over \sqrt{6} } \right)          \Bigr)^{-2}
\end{equation}
in Planck units. Here $\lambda_{\rm eff} $ depends on the composition of the inflaton.
For Higgs inflation, $\lambda_{\rm eff} =\lambda_h$; for singlet inflation,
$\lambda_{\rm eff} =\lambda_s /x^2$ with $x= \xi_s / \xi_h$; and for mixed inflation,
$\lambda_{\rm eff} = 
(4 \lambda_s \lambda_h - \lambda_{hs}^2)/\left[ 4( \lambda_s  + \lambda_h x^2 - \lambda_{hs} x  )  \right]   $.

Depending on the values  of $\lambda_i$ and $\xi_i$, 3 variants of inflation are possible.
Since the shape of the potential is the same in all cases (at tree level), 
they all share the same
prediction for the spectral index $n \simeq 1- 2/N $ and the tensor to scalar perturbation ratio 
$r \simeq 12/N^2$, with $N$ being the number of $e$-folds during inflation \cite{Bezrukov:2007ep}.  
Obviously, the prerequisite for inflation is that the relevant $\lambda_{\rm eff}$ be positive 
at high energies.
Then, given the low value of the Higgs mass, Higgs inflation becomes problematic 
\cite{Bezrukov:2008ej,DeSimone:2008ei,Barvinsky:2009ii}.  
To revive this option, one may use the mechanism of $\lambda_h$--enhancement described
in the previous section.

\subsection{Singlet--assisted  Higgs inflation  }

Although the minimalistic version of the original Higgs inflation  \cite{Bezrukov:2007ep}    
appears disfavored, the presence of an almost decoupled singlet can save the idea. 
Of course, the Higgs field is no longer special in this case as there is another scalar
which could drive inflation. Yet,  Higgs inflation remains an interesting possibility.

\begin{figure}
\hspace*{1.0cm}
\includegraphics[width=5.6cm]{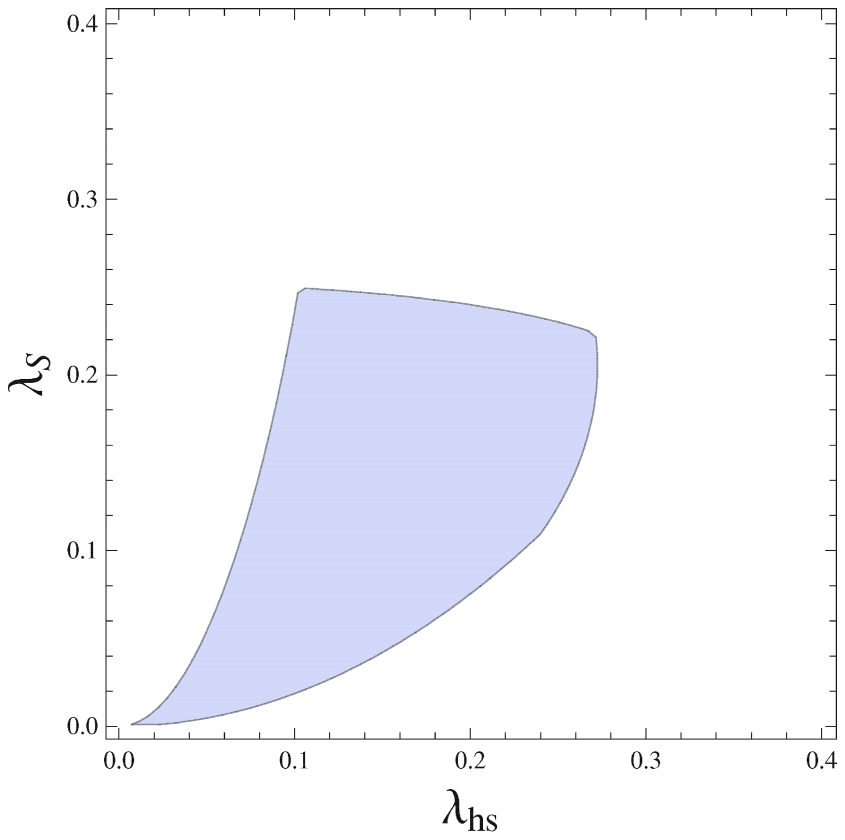} \hspace{0.9cm}
\includegraphics[width=5.6cm]{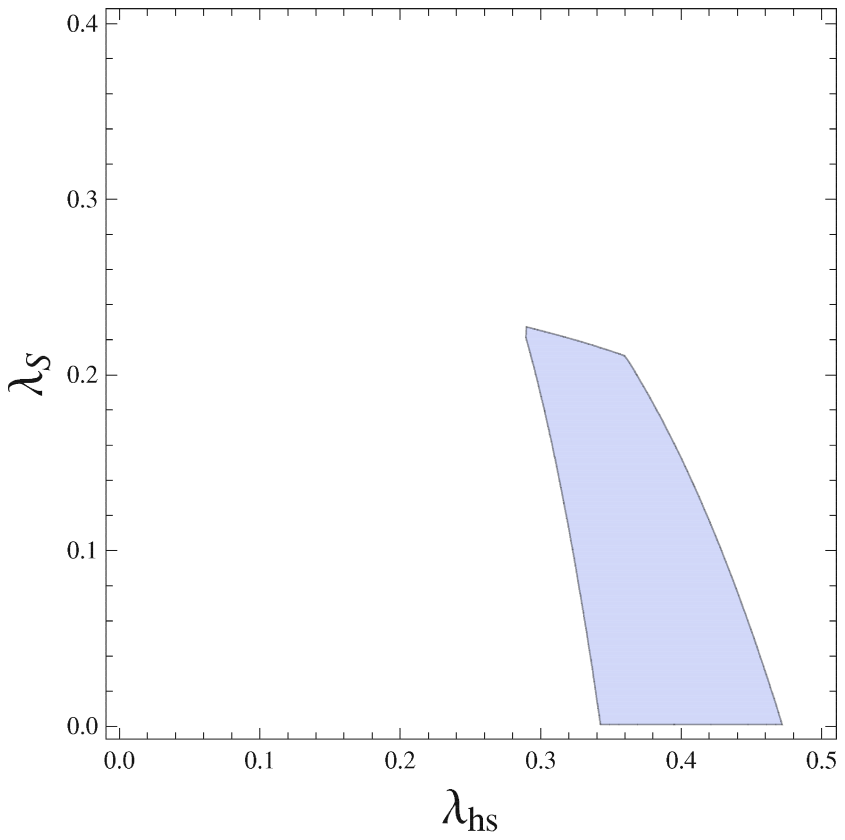}
\vspace*{0.3cm}
 \caption{\footnotesize  Parameter space allowed by Higgs inflation for
the two vacua: $u \gg v$ (left) and $u=0$ (right).  
   The Higgs mass is fixed at
approximately 125 GeV and $\xi_s/\xi_h = 10^{-3}$. }
\label{f2} 
\end{figure}

For Higgs inflation to occur, we require the following conditions
at the Hubble scale $\mu_H \sim M_{\rm Pl}/\xi_h$ with $\xi_h \sim  5\times 10^4$  
\cite{Bezrukov:2007ep} :
\begin{eqnarray}
&& \lambda_h >0 \;, \nonumber \\
&& 2 \lambda_h ~{\xi_s \over \xi_h} - \lambda_{hs} <0 \;. 
\end{eqnarray}
The second constraint follows from (\ref{taumin}) and ensures 
that $\tau=\infty$ is a local minimum. (The sign of 
$   2 \lambda_s \xi_h - \lambda_{hs} \xi_s  $ determines whether this 
is the only local minimum.) It follows then that  $\lambda_{hs}$ must
be positive, $\lambda_{hs} >0 $. 
Furthermore, we impose the ``perturbativity" constraint
\begin{equation}
 \vert \lambda_i \vert <1 \;
\end{equation}
at the Hubble scale. This judicious choice is motivated by perturbativity
($\lambda_i^2/(4 \pi)<1$)  all the way up to the Planck scale.

We take a conservative viewpoint  \cite{Lebedev:2011aq}  and do not impose further constraints
arising from loop corrections to the shape of the inflaton potential (the ``running'' spectral
index). These are  likely to be affected by the presence of higher dimensional operators
and/or heavy states \cite{Burgess:2009ea,Barbon:2009ya,Lerner:2010mq,Giudice:2010ka,Hertzberg:2011rc}. 
In the absence of a UV complete theory, such subtle  effects cannot 
be calculated reliably.

The allowed parameter space for $\xi_h \gg \xi_s$
is shown in Fig.~\ref{f2}. Here we set the singlet mass in the TeV 
range, so the  threshold effects are unimportant.
The left panel corresponds to the 
$u \gg v$ vacuum  such that 
$\lambda_h -\lambda_{hs}^2 / (4 \lambda_s)    \simeq 0.13$ at the EW scale. The right panel
corresponds to $u=0$ and  $\lambda_h \simeq 0.13  $ at the EW scale, in which case
the stabilization is due to loop effects. The obvious difference is that the first option allows
for arbitrarily small couplings $ \lambda_{hs},\lambda_s $, whereas in the other case
 $ \lambda_{hs}$ must be greater than 0.3-0.4. The allowed region is bounded on the right and from 
above by perturbativity, and on the left by $\lambda_h(\mu_H)>0$ in both cases. The constraint
 $2 \lambda_h ~{\xi_s / \xi_h} - \lambda_{hs} <0$ amounts essentially to positivity of $\lambda_{hs}$.
Finally, we note that the magnitude of $\xi_h \sim 5\times 10^4$ is fixed by the potential normalization 
\cite{Bezrukov:2007ep}.

The two possibilities can potentially be distinguished at colliders. The $u\gg v $ option can lead to 
 a significant correction to the Higgs self--coupling $\lambda_h$, whereas the $u=0$ case
requires a substantial Higgs coupling to the singlet. To observe production of the 
 (EW--scale)  singlet pairs
would be challenging, but feasible given very high luminosities \cite{Djouadi:2011aa}.

Finally, we note that other variants of Higgs portal inflation are  possible.
The inflaton can also be a mixture of the Higgs and the singlet as well as the singlet alone,
depending on $\lambda_i $ and $\xi_i$ \cite{Lebedev:2011aq}. To disentangle all the possibilities
at colliders would be very challenging as it requires determination of the sign of $\lambda_{hs}$.

\section{Conclusion}

We have studied  the possibility that the Higgs boson has a small admixture of an  SM singlet.
We find that, as the singlet VEV increases, 
the Higgs mass--coupling relation recieves a tree level contribution which does not vanish
in the  zero--mixing/heavy--singlet limit.  Such a correction can be order one and make 
the EW vacuum completely stable rather than metastable. 
The requisite Higgs--singlet coupling $\lambda_{hs}$  
and the singlet self--coupling $\lambda_s$   are allowed to be
very small, 
as long as $\lambda_{hs}^2/(4\lambda_s)$ is greater than about $0.015$ for a TeV--scale singlet.
This situation is practically indistinguishable from the SM at low
energies unless the Higgs quartic coupling is measured.

We also find that Higgs inflation is possible in our framework since the quartic coupling
remains positive at high energies.   This result can also be achieved purely through
singlet--induced loop effects, although a substantial Higgs--singlet coupling would be 
required in this case.

{\bf Acknowledgements.} The author is grateful to H.M. Lee for useful comments.

\end{document}